\documentclass[twocolumn]{pasj00}
\SetRunningHead{S. Nakagawa}{Multi-Color Photometry of the Outburst of
OT~J012059.6+325545}
\Received{2012 January 28}
\Accepted{2013 February 16}
\begin{document}

\title{Multi-Color Photometry of the Outburst of the New WZ~Sge-Type
Dwarf Nova, OT~J012059.6+325545}

\author{Shinichi \textsc{Nakagawa}, Ryo \textsc{Noguchi},
Eriko \textsc{Iino}, Kazuyuki \textsc{Ogura}, Katsura \textsc{Matsumoto}}
\affil{Osaka Kyoiku University, Asahigaoka 4-698-1, Kashiwara, Osaka 582-8582}
\email{katsura@cc.osaka-kyoiku.ac.jp (KM)}

\author{Akira \textsc{Arai}\thanks{Present address: Nishi-Harima
Astronomical Observatory, Center for Astronomy, University of Hyogo,
407-2, Nishigaichi, Sayo-cho, Sayo, Hyogo, 679-5313},
Mizuki \textsc{Isogai}}
\affil{Koyama Astronomical Observatory, Koyto Sangyo University,
Motoyama, Kamigamo, Kita-ku, Kyoto, 603-8555}

\author{Makoto \textsc{Uemura}}
\affil{Hiroshima Astrophysical Science Center, Hiroshima University,
Kagamiyama 1-3-1, Higashi-Hiroshima, Hiroshima 739-8526}


\maketitle

\begin{abstract}
We present our photometric studies of the newly discovered optical
transient, OT J012059.6+325545, which underwent a large outburst
between 2010 November and 2011 January.
The amplitude of the outburst was about 8~mag.
We performed simultaneous multi-color photometry by using $g^\prime$,
$R_{\rm C}$, and $i^\prime$-band filters from the early stage of the
outburst. 
The time resolved photometry during the early stage revealed periodic
variations with double-peaked profiles, which are referred to as
early superhumps, with amplitudes of about 0.08~mag.
After the rapid fading from the main outburst, we found
rebrightening phenomena, which occurred at least nine times.
The large amplitude of the outburst, early superhumps, and
rebrightening phenomena are typical features of WZ Sge-type dwarf novae.
We detected color variations within the early superhump modulations
making this only the second system, after V445 And, for which this has
been established.
We carried out numerical calculations of the accretion disk to explain
both of the modulations and the color variations of the early superhump.
This modeling of the disk height supports the idea that height
variations within the outer disk can produce the early superhump
modulations, though we cannot rule out that temperature asymmetries may
also play a role.
\end{abstract}

\section{Introduction}

Dwarf novae are a subclass of cataclysmic variables that consist
of a white dwarf with an accretion disk and a late-type main-sequence
(secondary) star donating its material to the accretion disk (Warner
1995 for a review).
Dwarf novae show sudden increases of their brightness which are called 
(normal) outbursts.
The outburst is currently understood to be triggered by a thermal
instability of the accretion disk (e.g., H$\bar{\rm o}$shi 1979).

SU UMa-type dwarf novae are one of the subgroups of dwarf novae with
orbital periods shorter than 2 hours, and show superoutbursts.
Superoutbursts are caused by tidal instability of the disk
(Whitehurst 1988; Osaki 1989), and are different from normal
outbursts in observed features and their mechanisms.
Superoutbursts show larger amplitudes and longer durations (lasting
for more than 10 days) than normal outbursts. 
The thermal-tidal instability model expects that a tidal instability
occurs when the disk expands beyond the 3:1 resonance radius
during a superoutburst (Osaki 1996 for a review), then the disk
should be deformed to an eccentric form, and undergoes a slow
precession (Vogt 1982; Osaki 1985; Whitehurst \& King 1991).
This acts as a trigger of superhumps which are short term modulations
with amplitudes of about 0.3--0.4 mag, and the period is a few percent
longer than the orbital period (Vogt 1974; Warner 1975).

WZ Sge-type dwarf novae are an extreme subgroup of SU UMa-type stars.
The interval of two successive superoutbursts (supercycle) can be
many years up to decades (O'Donoghue et al.\ 1991; Osaki 1995).
SU UMa-type stars usually exhibit several normal outbursts during one
supercycle, but WZ Sge-type stars show superoutbursts only and their
amplitudes are significantly larger (6--8~mag) than those of SU UMa-type
stars.
Furthermore, WZ Sge-type stars present two unique behaviors remaining to
be understood; early superhumps and post-superoutburst rebrightenings.
These peculiarities of WZ Sge-type stars have not been well understood
due to the low frequency of their superoutbursts.

Early superhumps are observed in very early phases of superoutbursts
before ordinary superhumps emerge, and the period of early superhumps is
almost in agreement with the orbital period (e.g., Kato et al.\
1996; Ishioka et al.\ 2002; Osaki \& Meyer 2002; Patterson et al.\ 2002).
A candidate that causes early superhumps is two-armed spiral pattern
excited by tidal dissipation which is generated by the 2:1 resonance
(Osaki \& Meyer 2002).
Other theoretical models of early superhumps based on tidal distortion
of the disk have been proposed (Smak 2001; Kato 2002; Ogilvie 2002).
Maehara et al.\ (2002) succeeded in reproducing the light curves of the
early superhumps of BC UMa, assuming spiral shock patterns on the disk.
Some SPH simulations demonstrated the appearance of the two-armed
spiral feature at the 2:1 resonance radius of the disk
(Kunze 2004; Kunze \& Speith 2005).
In contrast, observational evidences for these models have not been
established yet.
Some spiral or arch-like structures have been observationally detected
in WZ Sge-type stars via Doppler and eclipse mapping techniques, and
physical explanation of early superhumps is under discussion
(see e.g., Steeghs 2001; Smak 2001; Kato 2002 and references therein).
Recently Matsui et al.\ (2009) detected the early superhumps got
redder at maxima in simultaneous multi-color observations of V455 And,
and argued that the light source of the early superhump is probably
a low-temperature and vertically expanded region at the outermost
part of the disk.

Rebrightenings are additional brightening phenomena seen after rapid
fadings from superoutbursts of WZ Sge-type stars (Kato et al.\ 2004).
The rebrightenings can be classified by the shape of the light curve
into three types, (1) single short rebrightening, (2) repetitive short
rebrightenings, and (3) long lived plateau (Imada et al.\ 2006).
The thermal-tidal instability model is difficult to explain the
rebrightening at present, since most of gas in the disk should be
consumed during superoutbursts (Osaki 1996).
Kato et al.\ (1998) supposed that the gas which remain in the outer
regions beyond the 3:1 resonance radius could be transferred a few days
after the end of the main outburst, and triggering the rebrightenings.

In this paper we report the results of our observation of a new optical
transient OT J012059.6+325545 (or sometimes OT J012059.59+325545.0;
hereafter referred to as OTJ0120).
A bright outburst of the object was discovered at 12.3 mag by
K.~Itagaki on 2011 November 30.50663 (vsnet-alert 12431\footnote{\tt http://ooruri.kusastro.kyoto-u.ac.jp/mailarchive/vsnet-alert/12431}).
We carried out simultaneous $g^\prime$, $R_{\rm C}$, and $i^\prime$-band
multi-color photometry, from the day following the outburst detection to
the rebrightening phase.
There is a quiescent counterpart of OTJ0120 in the Sloan Digital Sky
Survey DR8 database with magnitudes of $g = 20.09$, $r = 20.24$
and $i = 20.52$.
Hence the amplitude of the outburst was $\sim 8$ mag.
Based on this large amplitude, OTJ0120 was considered as a new candidate
of WZ Sge-type dwarf nova.

\section{Observations}

Simultaneous multi-band optical observations were performed with
ADLER (Araki telescope DuaL-band imagER) attached to the 1.3-m ARAKI
telescope at Koyama Astronomical Observatory (KAO), and with an Andor
DW436 CCD camera attached to the 51-cm telescope at Osaka Kyoiku
University (OKU).
ADLER is an imager which can take images simultaneously in two
optical bands.
We obtained time-series data that allowed us to study short-term
variations, using a simultaneous photometry mode of ADLER for
$g^\prime$ and $i^\prime$-band observations and the Andor camera
for $R_{\rm C}$-band observations.
The date of observations covered from 2010 December 1 to 2011 January 8
at KAO and from 2010 December 1 to 2011 January 19 at OKU.
The exposure times were 15--180~s for the $g^\prime$ and
$i^\prime$-band observations and were 10--180~s for 
the $R_{\rm C}$-band observations, respectively, and
we also obtained calibration frames.
We measured the magnitudes of the object and comparison stars,
using APPHOT package of IRAF (Image Reduction and Analysis Facility)
for the $R_{\rm C}$-band data.
We also performed standard aperture photometry for the $g^\prime$
and $i^\prime$-band data using DAOPHOT package of IRAF in a software
developed by MI for the KAO data reduction pipeline.
As comparison stars, we used NOMAD 1229-0022689 for $g^\prime$
and $i^\prime$-band observations, and used NOMAD 1229-0022670
($B=12.529$, $V=12.577$, $R=12.610$) for the $R_{\rm C}$-band
observations during the main outburst phase and then
NOMAD 1229-0022691 ($B=16.260$, $V=16.130$, $R=16.460$) for
the post outburst phase.
Since there were no measurements of $g^\prime$ and $i^\prime$-magnitudes
of NOMAD 1229-0022689 at that time, we performed absolute photometry
with $g^{\prime}$ and $i^{\prime}$-bands using standard stars on a fair
night and corrected for the airmass by observing standard stars at
different altitudes.
As a result we estimated $g^\prime$ and $i^\prime$-magnitudes of
NOMAD 1229-0022689 as $g^\prime=14.256 \pm 0.060$ and
$i^\prime=13.648 \pm 0.056$.
We measured constancy of the comparison stars using other local stars,
and confirmed no significant variation in the brightness of the
comparison stars during our observations.

\section{Results}

\subsection{Light Curve Analysis}

\begin{figure}
 \begin{center}
  \FigureFile(80mm,50mm){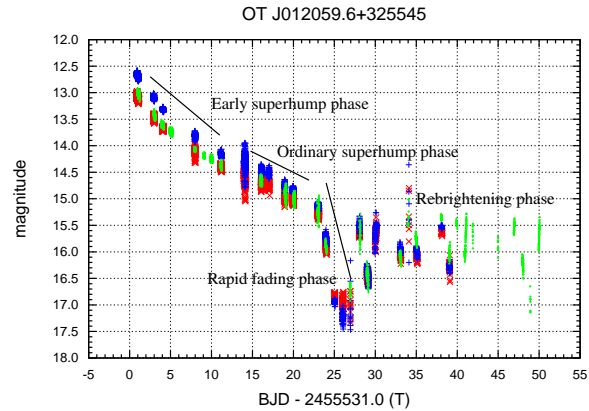}
 \end{center}
 \caption{ $g^\prime$, $R_{\rm C}$ and $i^\prime$-band light curves of
 the 2010 outburst of OTJ0120.
 The horizontal axis displays time in days from BJD = 2455531.0 (the
 day of the first detection of this outburst).
 The blue pluses, green points and red crosses represent $g^\prime$,
 $R_{\rm C}$ and $i^\prime$-band observations,
 respectively.}\label{fig:lc}
\end{figure}

Figure~\ref{fig:lc} shows the light curve with all the photometric
data we obtained by the observations.
In this paper, we denote the number of days elapsed from the detection
of the outburst as $T$, and $T=0$ corresponds to JD = 2455531.
We divided the data into four phases by the features of the light curve,
i.e., the early superhump phase (from $T=1$ to $T=10$), the ordinary
superhump phase (from $T=11$ to $T=20$), the rapidly fading phase (from
$T=23$ to $T=27$), and the rebrightening phase (from $T=28$ to $T=50$).
After our first observation on $T=1$, the flux decreased from
$R_{\rm C}=13.0$ to 15.3 for 23 days with a slow declining rate of
about 0.10~${\rm mag~d^{-1}}$.

We examined the rate of decline during both the early superhump phase
and the ordinary superhump phase in the $R_{\rm C}$-band were
0.14~${\rm mag~d^{-1}}$ and 0.09~${\rm mag~d^{-1}}$, respectively.
The large decline rate in the early superhump phase is a typical feature
of superoutbursts of WZ Sge-type stars (e.g., Nogami et al.\ 1997).
We then detected a rapid fading from the outburst on $T=24$.
The rapid fading stage continued at a decline rate of
0.65~${\rm mag~d^{-1}}$, examined in the $R_{\rm C}$-band
until $T=25$ when the object was $g^\prime=16.8$.
We made the first detection of a dramatic rebrightening
that increased the flux to $g^\prime=15.5$ at $T\sim28$.
In total, we detected nine rebrightenings as marked by arrows
in Figure~\ref{fig:reb}, while additional rebrightenings might
have been overlooked in our observations.
The average rising and fading rates of the rebrightenings were
3.3~${\rm mag~d^{-1}}$ and 0.95~${\rm mag~d^{-1}}$
examined in the $R_{\rm C}$-band, respectively.
We obtained the longest coverage of a rising trend on $T=41$, which
indicated $-0.59$~mag in 0.14~d as a lower (shorter) limit of
the rise time or timescale $\tau \sim 0.25$~d when the time series
of flux $f(t)$ was supposed as $f(t) = \exp(t/\tau) \times C$.
As for the decay time, we similarly obtained a lower limit of 
$+0.23$~mag in 0.26~d on $T=29$ which was represented
by $\tau \sim -1.1$~d.

\begin{figure}
 \begin{center}
  \FigureFile(80mm,50mm){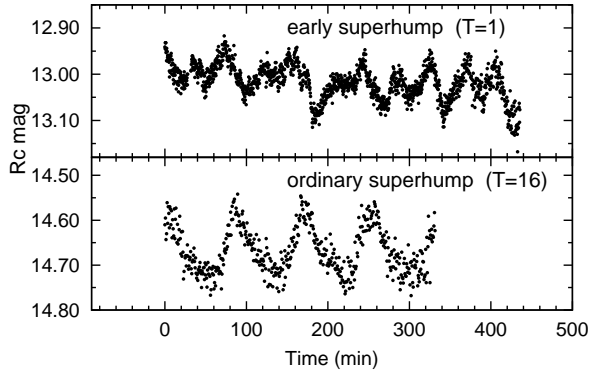}
 \end{center}
 \caption{Samples of the short term modulations observed in the 2010
 superoutburst of OTJ0120.
 The upper panel shows the double-peaked modulations detected on $T=1$.
 The lower panel shows the single-peaked modulations detected on
 $T=16$.}\label{fig:shortlc}
\end{figure}

\begin{figure}
 \begin{center}
  \FigureFile(80mm,50mm){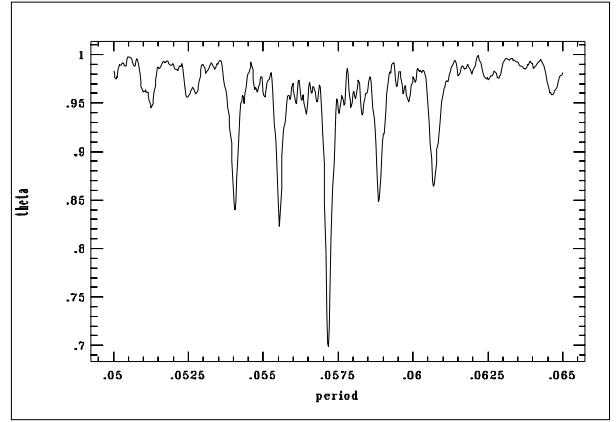}
 \end{center}
 \caption{Period-theta diagram obtained from the PDM analysis for
 double-peaked modulations.}\label{fig:earlypdm}
\end{figure}

\begin{figure}
 \begin{center}
  \FigureFile(80mm,50mm){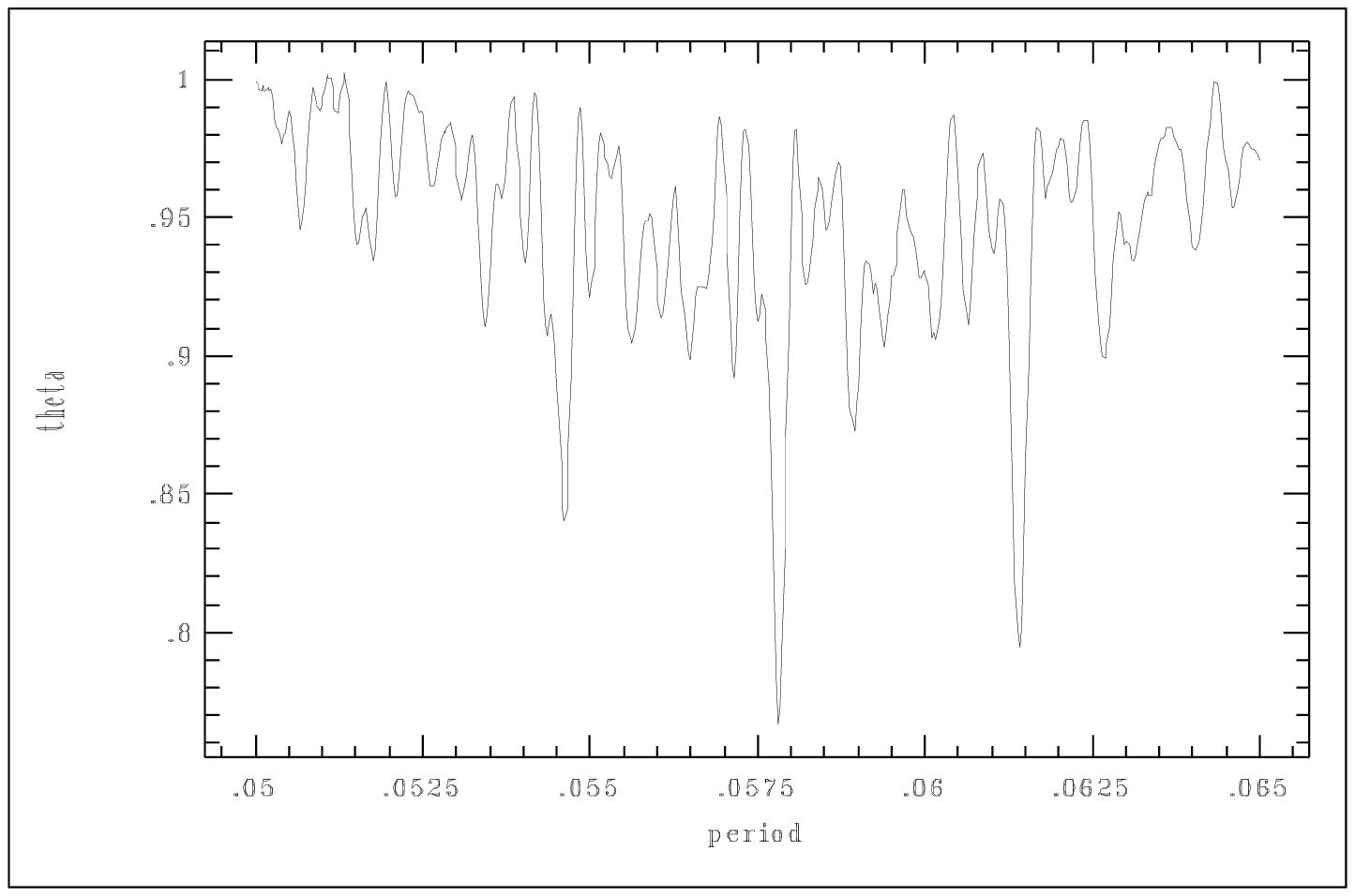}
 \end{center}
 \caption{Period-theta diagram obtained from the PDM analysis for
 single-peaked modulations.}\label{fig:SHpdm}
\end{figure}

We found double-peaked modulations during $T=1$--$10$ as shown
in the upper panel of Figure~\ref{fig:shortlc}.
Subsequently, the modulations decayed and were replaced by
single-peaked modulations which were detected during $T=11$--$23$.
We estimated periods of the two types of modulations by the Phase
Dispersion Minimization (PDM) method (Stellingwerf 1978) using
the time-series data for which global declining trends were subtracted.
The resulting period-theta diagrams are shown in 
Figure~\ref{fig:earlypdm} and Figure~\ref{fig:SHpdm}.
The errors were estimated by the Lafler-Kinman method (Fernie 1989).
We used the data during $T=1$--$10$ for Figure~\ref{fig:earlypdm}
and $T=11$--$23$ for Figure~\ref{fig:SHpdm}.
The best-estimated period of the double-peaked modulations and the
single-peaked modulations are 0.057145 $\pm$ 0.000002~d and
0.057814 $\pm$ 0.000012~d, respectively.
The period of the double-peaked modulations were $1.17 \pm 0.02\%$,
shorter than that of the single-peaked modulations,
and was stable across the duration.
Our analysis indicates that the double-peaked modulations observed
during the early phase were early superhumps, and the single-peaked
modulations observed during the late phase were ordinary superhumps.
We thus conclude that OTJ0120 is a WZ Sge-type dwarf nova because of
the existence of the early superhumps and the rebrightenings as well as
the large amplitude of the outburst.

\subsection{Superhump Period Change}

\begin{figure}
 \begin{center}
  \FigureFile(120mm,150mm){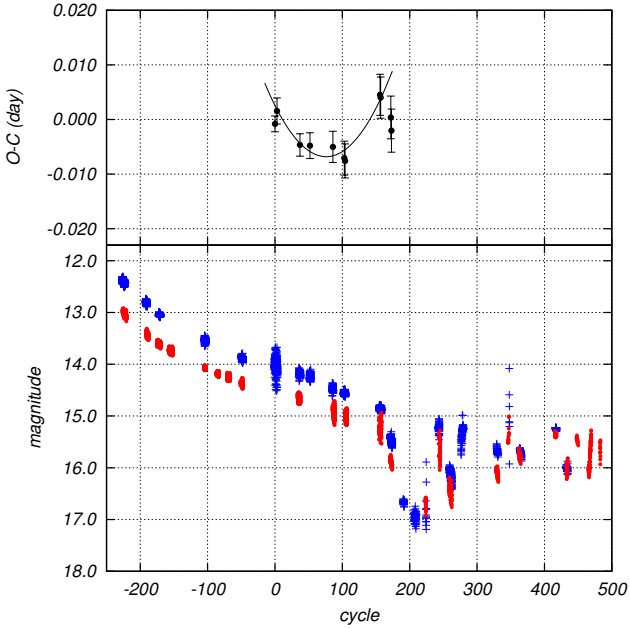}
 \end{center}
 \caption{$g^\prime$ and $R_{\rm C}$-band light curves and the $O-C$
 versus cycle count (E). 
 Upper panel: $O-C$ versus cycle count (E) of superhumps.
 Lower panel: $g^\prime$ and $R_{\rm C}$-band light curves.
 The blue pluses and red points represent $g^\prime$ and
 $R_{\rm C}$-band observations, respectively.}\label{fig:g+OC}
\end{figure}

We searched for period changes of the ordinary superhumps during the
superoutburst by the following method.
A template light curve of the superhump was defined as a
phase-averaged light curve of the superhumps obtained during the
ordinary superhump phase.
We then shifted the maximum timing of the template light curve
around an eye-estimated maximum of each observation
by 0.0001~d steps and calculated variances of each test.
Finally the timings of the maxima were determined by minimizing
the variances, which are given in Table\ref{tab:O-C}.
The cycle count ($E$) is defined as $E=0$ at the first superhump
maximum we observed.
A linear ephemeris of the superhump maxima is given by
$BJD(maximum) = 2455544.9890(3) + 0.057814(12)E$.
The resulting $O-C$ diagram is shown in Figure~\ref{fig:g+OC}.
We can see a gradual decreasing trend in $O-C$ until $E=104$ and a
subsequent increasing trend, which indicate an increase of the superhump
period during the superoutburst.

\begin{table}
 \begin{center}
  \caption{Times of superhump maxima.}\label{tab:O-C}
  \scalebox{0.8}{
  \begin{tabular}{lccc}
   \hline
   \multicolumn{1}{c}{BJD$^{*} - 2455500$} & Error ($10^{-3}$) & E(cycle) & $O-C$ ($10^{-3}$ day) \\
   \hline
   44.98824 & 1.4 & 0 & $-0.8$ \\
   45.16403 & 2.4 & 3 & 1.6 \\
   47.12334 & 2.1 & 37 & $-4.7$ \\
   47.99036 & 2.4 & 52 & $-4.8$ \\
   49.95567 & 2.8 & 86 & $-5.0$ \\
   50.93637 & 3.1 & 103 & $-7.1$ \\
   50.99367 & 3.1 & 104 & $-7.6$ \\
   54.01186 & 3.8 & 156 & 4.5 \\
   54.06916 & 3.8 & 157 & 4.0 \\
   54.93267 & 3.9 & 172 & 0.4 \\
   54.98806 & 3.9 & 173 & $-2.0$ \\
   \hline
   \multicolumn{4}{l}{{\small$^{*}$Barycentric Julian Date (Eastman,
   Siverd \&Gaudi 2010)}} \\
  \end{tabular}
  }
 \end{center}
\end{table}

\subsection{Phase-Averaged Light Curve and Color of Early Superhumps}

\begin{figure}
 \begin{center}
  \FigureFile(80mm,50mm){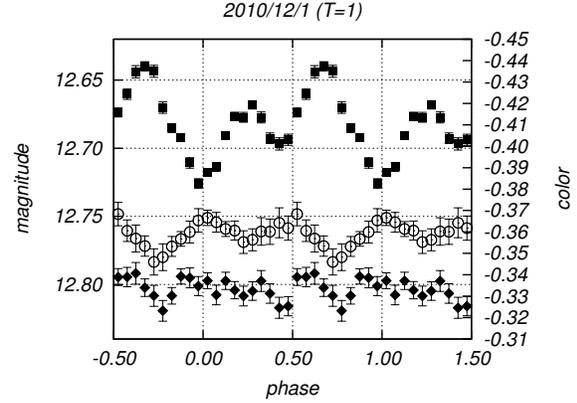}
 \end{center}
 \caption{Phase averaged light curve on $T=1$.
 The squares represent the $g^\prime$-band observations.
 The diamonds represent $g^\prime-R_{\rm C}$ color and
 the circles represent $g^\prime-i^\prime$ color plus 0.05
 offset.}\label{fig:esh}
\end{figure}

We searched for short-term color variations in the early superhumps.
The phase-averaged $g^\prime$-band light curve and the color
($g^\prime-R_{\rm C}$ and $g^\prime-i^\prime$) on $T=1$ are shown in
Figure~\ref{fig:esh}.
We found the colors of $g^\prime-R_{\rm C}$ and $g^\prime-i^\prime$
were largest, i.e.\ reddened, at phase $\sim$ 0.7 which corresponds to
the maximum of the $g^\prime$-band light curve.
In addition, the color minima were almost coincident with the
minimum of the $g^\prime$-band light curve.
The same trend of color variations was also seen in the data
on $T=3$ and $T=4$ (we did not use the data at $T=2$ and between $T=5$
and $7$ in this analysis because of the short duration and the absence
of simultaneous multi-color observations, respectively).
These facts indicate that OTJ0120 is the second example of dwarf novae
for which color variations of early superhumps are determined,
and we confirmed that the two examples show the same trend for colors
to be redder at maxima.

The amplitudes of the early superhumps, which declined day by day,
were about 0.08~mag on $T=1$, 0.06~mag on $T=3$, 0.04~mag on $T=4$,
and 0.04~mag on $T=8$ in the $g^\prime$-band light curve.
We detected larger humps on $T=11$ (amplitude $\sim 0.08$~mag),
and the period of the modulations was different from that of the
early superhumps.
Therefore we interpret the growth of ordinary superhumps started around
$T=11$.

\section{Discussions}

\subsection{The Period of the Short Term Modulations}

The superhump period of OTJ0120 is significantly shorter than those of
general SU UMa-type dwarf novae, and is typical for WZ Sge-type dwarf
novae.
The estimation of superhump period excess is important for the study of
dwarf novae, because it gives us a way to estimate the mass ratio, and
thereby to investigate the evolutionary status of objects.
WZ Sge-type dwarf novae are especially predominant in such studies
because those orbital periods are close to the suspected period
minimum of CVs, so that the phenomenology of WZ Sge-type outburst
behavior is a vital example in studies of evolutionary states of CVs.
However, no direct measurement of the orbital period of OTJ0120 has been
performed; no eclipsing phenomenon has been detected, and no periodic
variation has been observed in quiescence.
Thus, we calculated the superhump period excess from the early
superhump period, which is likely equal to the orbital period
(Patterson et al.\ 1981; Kato et al.\ 1996).
The superhump period excess is
$\epsilon = P_{SH}/P_{orb}-1 =
  P_{SH}/P_{ESH}-1 = 1.17 \pm 0.02 \times 10^{-2}$,
which may be a lower limit if there is some level of precession
in the disk.
According to Kato et al.\ (2009), the empirical law exists between
the superhump period excess and mass ratio.
We derived the mass ratio $q = (M_1/M_2) = 15.2$ from the
superhump period excess which was estimated by our observations. 
The extreme mass ratio and the inferred short orbital period are
strongly in line with evolved CVs such as WZ Sge, which reinforces
the principle that WZ Sge-type dwarf novae, including OTJ0120, are
evolved CVs which are close to or passed beyond the minimum orbital 
period of CVs (sometimes called ``period bouncers''; Patterson 1998).

Olech et al.\ (2003) and Soejima et al.\ (2009) found that short
$P_{SH}$ systems showed period changes of superhumps with three
stages in $O-C$ diagrams.
Kato et al.\ (2009) reported general characteristics of the superhump
period changes, based on observations of superoutbursts of 199 dwarf
novae.
They discovered that the period change in superhump periods can be
categorized into three stages: (A) an early stage of superhump evolution
having a longer $P_{SH}$, (B) a middle segment with a stabilized period,
usually with a positive $\dot{P}$, and (C) a late stage with a shorter
stable superhump period.
Figure~\ref{fig:g+OC} shows the $g^\prime$ and $R_{\rm C}$-band light
curves and the $O-C$ versus cycle count ($E$) during the 2010
superoutburst of OTJ0120.
In general, the $O-C$ diagram of the superhumps is abruptly varied at
the timing of entering to a rapid fading phase among WZ Sge-type stars
(see figure~33 of Kato et al.\ 2009).
The $O-C$ variation of OTJ0120 can be interpreted as a transition from
stage B to C; the object was in a stage B during the superoutburst
between $E=3$ and $157$, in which the period derivative was positive.
Then an abrupt decline of $O-C$ was seen in $E=172$ and $173$
at the beginning of the rapid fading phase.
This abrupt decline could be a sign of a transition to stage C.
Hence, the characteristics of the $O-C$ of superhump during the 2010
superoutburst of OTJ0120 was consistent with those of other WZ Sge-type
stars.
We estimated $P_{\rm dot}=\dot{P}/P = (2.7 \pm 0.4) \times 10^{-5}$
during the stage B.

\subsection{Rebrightening}

\begin{figure}
 \begin{center}
  \FigureFile(80mm,50mm){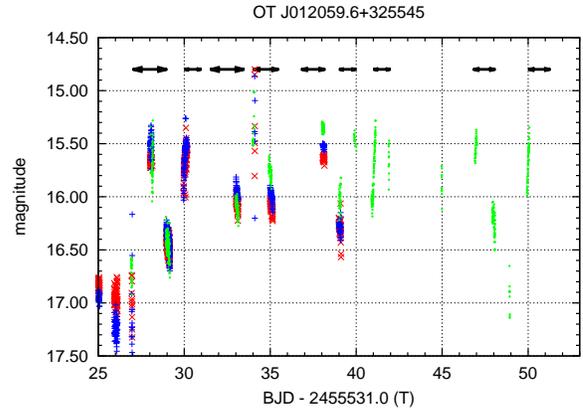}
 \end{center}
 \caption{Magnified view of the light curve around the rebrightening
 phase. The symbols are the same as those in Figure~1.}\label{fig:reb}
\end{figure}

Figure~\ref{fig:reb} shows a magnified view of the light curve around
the rebrightening phase.
We can see rebrightenings more than nine times and those amplitudes
were 1--2~mag.
The whole feature of the light curve in this phase is similar to that of
the 2001 outburst of WZ Sge (Ishioka et al.\ 2002).
The minimum following the first rebrightening was fainter than the
later ones.
Such a deep minimum can be also seen in the last dip.
Those characteristics were also analogous to the 2001 outburst of WZ Sge.
It is possible that we missed other deep dip, taking the durations
of our observations into account.

\begin{figure}
 \begin{center}
  \FigureFile(100mm,60mm){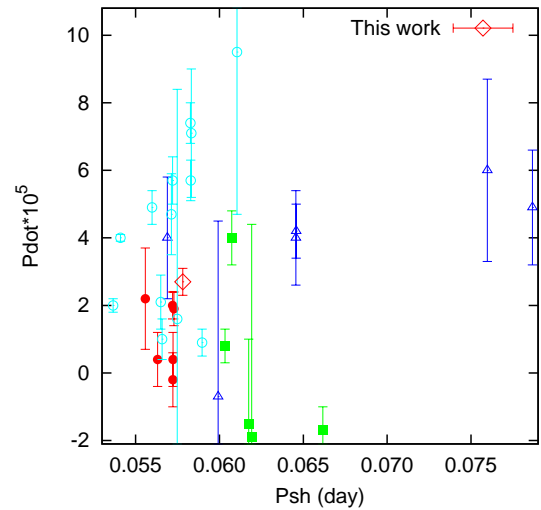}
 \end{center}
 \caption{Distribution of superhump period derivative for each type of
 rebrightenings.
 The diamond represents our result.
 The other data are from Kato et al.\ (2009).
 The symbols represent types of rebrightenings; type-A (filled circles),
 type-B (filled squares), type-C (open triangles) and type-D (open
 circles).}\label{fig:kato2009_reb}
\end{figure}

Kato et al.\ (2009) discussed the relation between $P_{\rm dot}$ and
light curve shapes of rebrightenings of WZ Sge-type stars.
They classified the rebrightenings into four types by their light curve
shapes; long duration rebrightenings (type-A), multiple rebrightenings
(type-B), single rebrightening (type-C), no rebrightening (type-D; see
figure~37 of Kato et al.\ 2009). 
The 2010 superoutburst of OTJ0120 is considered to be a type-A
outburst based on the classification.
Figure~\ref{fig:kato2009_reb} shows $P_{\rm dot}$ against $P_{\rm SH}$.
The figure includes the data reported in Kato et al.\ (2009) and
our result of OTJ0120.
The filled circles, filled squares, open triangles and open circles
represent type-A, type-B, type-C, and type-D outbursts, respectively.
Our result is expressed by an open diamond.
OTJ0120 has $P_{\rm dot}$ close to those of known type-A objects.
It supports that type-A objects tend to have smaller $P_{\rm SH}$ and
smaller $P_{\rm dot}$ among WZ Sge stars.
OTJ0120 had the largest $P_{\rm SH}$ and showed the largest
$P_{\rm dot}$ in the type-A objects.

\subsection{Color Variations of Early Superhumps}

Early superhumps are considered to be caused by a vertically expanded
disk which is, for example, explained by scenarios proposed by Kato
(2002) or Osaki \& Meyer (2002).
Matsui et al.\ (2009) discovered that the light source of
the early superhump of V455 And is connected with a low-temperature
component.
Our observation confirmed that OTJ0120 also became redder at both of
the primary and secondary maxima of the early superhumps.
This supports the argument that the light sources of the early
superhumps are expanded low-temperature components in outer regions
of the disk.

\begin{table}
 \begin{center}
  \caption{Model parameters for the early superhump mapping.}\label{tab:par}
  \begin{tabular}{lccc}
   \hline
   \multicolumn{1}{c}{Parameter} & Value \\
   \hline
   $\sigma {\rm (mag)}$          & 0.01 \\
   $R_{out}/a$                   & 0.58 \\
   $R_{in}/a$                    & 0.023 \\
   $T_{in} {\rm (K)}$            & 223500 \\
   Inclination angle (deg)       & 60.0 \\
   Mass ratio ($M_1/M_2$)        & 15.2 \\
   \hline
  \end{tabular}
 \end{center}
\end{table}

\begin{figure}
 \begin{center}
  \FigureFile(80mm,50mm){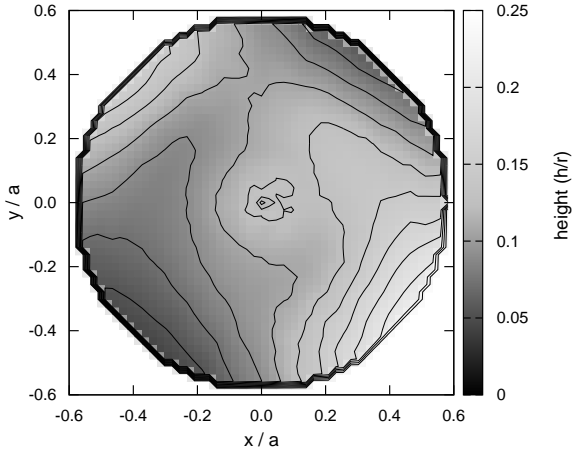}
 \end{center}
 \caption{The height map of the disk calculated from the data
 obtained on December 1 ($T=1$).
 The horizontal (x) and vertical (y) axes are represented by a unit of
 binary separation (a).
 The horizontal axis directs the center of the secondary, and 
 the vertical axis directs the motion of the secondary.
 The shade represents disk height, i.e., brighter regions are higher.
 The scale bar at the right side is represented by a unit of disk radius (r).
 The secondary star is located at $(x,y) = (1.0,0.0)$.}\label{fig:map}
\end{figure}

\begin{figure}
 \begin{center}
  \FigureFile(60mm,40mm){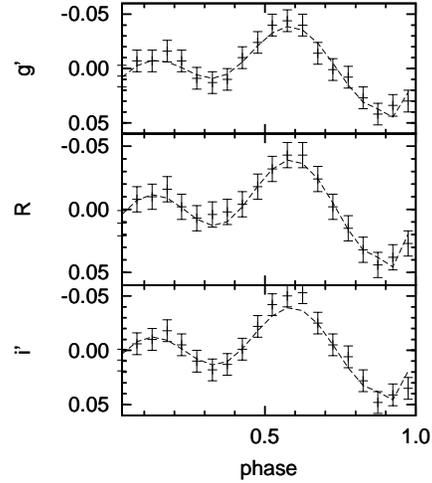}
 \end{center}
 \caption{Comparison of the light curve reproduced from the disk map
 (dashed lines) and the observed data (bars).
 The horizontal axis displays the orbital phase, and vertical axis
 displays the relative magnitudes of the $g^\prime$, $R_{\rm C}$, and
 $i^\prime$-bands.
 We defined the zero points of magnitudes as the averages in each
 band.}\label{fig:modellc} 
\end{figure}

We performed numerical calculations that allowed us to estimate vertical
structure of the disk assuming that the early superhump was caused by
the vertical deformations of the disk.
The early superhump mapping, which was developed by one of the authors
(MU), reconstructs disk structures using the Markov chain Monte Carlo
method based on the Bayesian statistics (Uemura et al.\ 2012).
The selected parameters for our calculations are summarized in
Table\ref{tab:par}, and we assumed the orbital period equals
the early superhump period.
We used $q = (M_1/M_2) = 15.2$, as discussed in subsection 4.1.
Assuming $0.6M_{\odot}$ as a typical mass of white dwarfs,
the radius ($R_1$) was $R_1 = 0.012 {\rm R_{\odot}}$ using
the mass-radius relation of white dwarfs (Provencal et al.\ 1998)
The binary separation was calculated as $a = 3.74 \times 10^{10}
{\rm cm}$ from the Keplerian law.
According to Osaki \& Meyer (2002), early superhumps should occur when
the disk radius enlarges and reaches to the 2:1 resonance radius.
We therefore set the outer radius of the disk $R_{out} = 0.6a$, which is
expected as the 2:1 resonance radius for the system.
We assumed that the inner radius of the disk was $R_1$, and supposed
that the inclination angle is lower than 65 degrees on the basis of the
absence of eclipses.
The lower limit of the inclination angle of OTJ0120 is poorly known.
Kato (2002) reported that the WZ Sge-type binaries with higher
inclination angles tended to indicate larger amplitudes of early
superhumps.
According to this trend, the detection of the early superhumps with an
amplitude of 0.08~mag indicates a moderate inclination angle of OTJ0120.
So we performed the calculation assuming an inclination angle of
$60^{\circ}$.
We estimated the disk temperature in the same way as Uemura et al.\ (2012).
The inner ($T_{in}$) and outer ($T_{out}$) temperatures of the disk were
estimated from the averaged color at each night.
The color index on $T=1$ was $g^\prime - i^\prime = -0.408$.
The best-fit temperatures for an inclination angle of $60^{\circ}$
were $T_{in} = 223500$~K and $T_{out} = 20000$~K.

Figure~\ref{fig:map} shows the disk map calculated for the inclination
angle of $60^\circ$.
Comparisons between the reproduced light curves from the calculated disk
and the observed ones are presented in Figure~\ref{fig:modellc}.
It shows that the reproduced light curves are almost in accordance with
the observational data.
We can see elevated structures in the lower right and the upper left
regions in Figure~\ref{fig:map}.
In addition, an armed structure extends from the lower-right elevated
area to the upper-right inner region.
These structures were seen even if we assumed other inclination angles
such as $40^\circ$ or $50^\circ$.
The calculations were based on an assumption of an axi-symmetric
distribution of temperature on the disk, which could actually be
axi-asymmetric due to the spiral wave, so that the armed structure
at smaller radii may be a high temperature region at larger radii.
The map reconstructed with the former hypothesis successfully
reproduced the observed light curves, and brought a similar pattern
of the disk to that of the theoretical models for tidally deformed
disk (e.g., Ogilvie 2002).
Such a distorted disk is expected to have a two-armed pattern,
while in our calculations there is no sign of the armed structure
which would be expected to appear in the lower-left region in
Figure~\ref{fig:map}.
It is therefore suggested that the tidal force would not be enough
to explain the structure of the disk leading to early superhumps.

\section{Summary}
We performed simultaneous multi-color photometry of the 2010 outburst
of OTJ0120.
The outburst exhibited a large amplitude ($\sim$ 8~mag), early
superhumps, and post-superoutburst rebrightenings.
The parameters obtained from period analyses, such as the superhump
period excess or $\dot{P}$, were consistent with those of other
WZ Sge-type stars discovered so far.
The rebrightenings were recorded nine times, and the whole light curve
was similar to that of the 2001 outburst of WZ Sge.
These results indicate that OTJ0120 is a new member of WZ Sge-type stars.
We found color variations of the early superhumps as a result of the
observations during the earliest phase of the outburst.
The phase averaged light curve of the early superhumps showed that
the humps were redder at maximum timings, and the characteristic of
the color variations was consistent with that of V455 And (Matsui et
al.\ 2009).
In order to explain such the modulations of the early superhumps,
we performed calculations to reconstruct the disk structure
by adopting a Bayesian model under some assumptions such as
an axi-symmetric temperature distribution on the disk.
The resulting disk maps indicated that the two vertically elevated
structures facing each other at the outer region of the disk could
reproduce the observed color variations, though we cannot rule out
that temperature asymmetries may also play a role.

\bigskip
This work was partly supported by the Private University Strategic
Research Foundation Support Program of the Ministry of Education,
Science, Sports and Culture of Japan (S0801061) and a Grand-in-Aid from
the Ministry of Education, Culture, Sports, Science, and Technology of
Japan (22540252).

\end{document}